%% file: main.tex
\documentclass[sigconf]{acmart}

\AtBeginDocument{%
  }

\copyrightyear{2023} 
\acmYear{2023} 
\setcopyright{acmlicensed}\acmConference[SIGMOD-Companion '23]{Companion of the 2023 International Conference on Management of Data}{June 18--23, 2023}{Seattle, WA, USA}
\acmBooktitle{Companion of the 2023 International Conference on Management of Data (SIGMOD-Companion '23), June 18--23, 2023, Seattle, WA, USA}
\acmPrice{15.00}
\acmDOI{10.1145/3555041.3589674}
\acmISBN{978-1-4503-9507-6/23/06}

\graphicspath{{img/}}
\usepackage[a-1b]{pdfx}
\usepackage{hyperref}
\usepackage{array}
\usepackage{amsmath}
\usepackage{algorithmic}
\usepackage[ruled,vlined]{algorithm2e}
\usepackage{pgfplots,pgfplotstable}  
\usepackage{booktabs}
\usepackage{balance}
\usepackage{multirow}
\usepackage{enumitem}
\usepackage{adjustbox}
\usepackage{subcaption}
\usepackage{tablefootnote}
\usepackage{fancybox}
\usepackage{soul}
\usepackage[most]{tcolorbox}
\usepackage{amsmath}
\usepackage{algorithmic}
\usepackage[ruled,vlined]{algorithm2e}
\usepackage{pgfplots,pgfplotstable}  
\usepackage{booktabs}
\usepackage{balance}
\usepackage{multirow}
\usepackage{enumitem}
\usepackage{adjustbox}
\usepackage{subcaption}
\usepackage{tablefootnote}
\usepackage{fancybox}
\usepackage{soul}
\newcommand{\edit}[1]{\textcolor{black}{#1}}

\newcommand{\mypar}[1]{~\noindent\textbf{#1.}\xspace}

\newcommand{\eat}[1]{}

\newcommand{\viewpoint}[2]{
\vspace{0.2cm}
\noindent\textbf{#1:} \textit{#2} 
\vspace{0.1cm}
}

\newcounter{enum}

\begin{document}

\settopmatter{authorsperrow=4}

\author{Yiwen Zhu}
\affiliation{
 \institution{Microsoft, USA}
}
\email{yiwzh@microsoft.com}

\author{Yuanyuan Tian}
\affiliation{
 \institution{Microsoft, USA}
}
\email{yuanyuantian@microsoft.com}

\author{Joyce Cahoon}
\affiliation{
 \institution{Microsoft, USA}
}
\email{jcahoon@microsoft.com}

\author{Subru Krishnan}
\affiliation{
 \institution{Microsoft, USA}
}
\email{subru@microsoft.com}

\author{Ankita Agarwal}
\affiliation{
 \institution{Microsoft, USA}
}
\email{ankiagar@microsoft.com}

\author{Rana Alotaibi}
\affiliation{
 \institution{Microsoft, USA}
}
\email{ranaalotaibi@microsoft.com}

\author{Jes\'us Camacho-Rodr\'iguez}
\affiliation{
 \institution{Microsoft, USA}
}
\email{jesusca@microsoft.com}

\author{Bibin Chundatt}
\affiliation{
 \institution{Microsoft, India}
}
\email{bibin.chundatt@microsoft.com}

\author{Andrew Chung}
\affiliation{
 \institution{Microsoft, China}
}
\email{andchung@microsoft.com}

\author{Niharika Dutta}
\affiliation{
 \institution{Microsoft, USA}
}
\email{nidutta@microsoft.com}

\author{Andrew Fogarty}
\affiliation{
 \institution{Microsoft, USA}
}
\email{anfog@microsoft.com}

\author{Anja Gruenheid}
\affiliation{
 \institution{Microsoft, USA}
}
\email{agruenheid@microsoft.com}

\author{Brandon Haynes}
\affiliation{
 \institution{Microsoft, USA}
}
\email{brhaynes@microsoft.com}

\author{Matteo Interlandi}
\affiliation{
 \institution{Microsoft, USA}
}
\email{mainterl@microsoft.com}

\author{Minu Iyer}
\affiliation{
 \institution{Microsoft, USA}
}
\email{minu.iyer@microsoft.com}

\author{Nick Jurgens}
\affiliation{
 \institution{Microsoft, USA}
}
\email{nicholas.jurgens@microsoft.com}

\author{Sumeet Khushalani}
\affiliation{
 \institution{Microsoft, USA}
}
\email{sukhusha@microsoft.com}

\author{Brian Kroth}
\affiliation{
 \institution{Microsoft, USA}
}
\email{bpkroth@microsoft.com}

\author{Manoj Kumar}
\affiliation{
 \institution{Microsoft, India}
}
\email{manok@microsoft.com}

\author{Jyoti Leeka}
\affiliation{
 \institution{Microsoft, USA}
}
\email{Jyoti.Leeka@microsoft.com}

\author{Sergiy Matusevych}
\affiliation{
 \institution{Microsoft, USA}
}
\email{sergiym@microsoft.com}

\author{Minni Mittal}
\affiliation{
 \institution{Microsoft, India}
}
\email{minni.mittal@microsoft.com}

\author{Andreas Mueller}
\affiliation{
 \institution{Microsoft, USA}
}
\email{amueller@microsoft.com}

\author{Kartheek Muthyala}
\affiliation{
 \institution{Microsoft, USA}
}
\email{kamuth@microsoft.com}

\author{Harsha Nagulapalli}
\affiliation{
 \institution{Microsoft, USA}
}
\email{hanagula@microsoft.com}

\author{Yoonjae Park}
\affiliation{
 \institution{Microsoft, USA}
}
\email{yoonjae.park@microsoft.com}

\author{Hiren Patel}
\affiliation{
 \institution{Microsoft, USA}
}
\email{hirenp@microsoft.com}

\author{Anna Pavlenko}
\affiliation{
 \institution{Microsoft, USA}
}
\email{anna.pavlenko@microsoft.com}

\author{Olga Poppe}
\affiliation{
 \institution{Microsoft, USA}
}
\email{olpoppe@microsoft.com}

\author{Santhosh Ravindran}
\affiliation{
 \institution{Microsoft, USA}
}
\email{santhosh.ravindran@}
\email{microsoft.com}

\author{Karla Saur}
\affiliation{
 \institution{Microsoft, USA}
}
\email{kasaur@microsoft.com}

\author{Rathijit Sen}
\affiliation{
 \institution{Microsoft, USA}
}
\email{rathijit.sen@microsoft.com}

\author{Steve Suh}
\affiliation{
 \institution{Microsoft, USA}
}
\email{stsuh@microsoft.com}

\author{Arijit Tarafdar}
\affiliation{
 \institution{Microsoft, USA}
}
\email{arijitt@microsoft.com}

\author{Kunal Waghray}
\affiliation{
 \institution{Microsoft, USA}
}
\email{kunalwaghray@microsoft.com}

\author{Demin Wang}
\affiliation{
 \institution{Microsoft, USA}
}
\email{deminw@microsoft.com}

\author{Carlo Curino}
\affiliation{
 \institution{Microsoft, USA}
}
\email{ccurino@microsoft.com}

\author{Raghu Ramakrishnan}
\affiliation{
 \institution{Microsoft, USA}
}
\email{raghu@microsoft.com}


\renewcommand{\shortauthors}{Yiwen Zhu et al.} 

\title{Towards Building Autonomous Data Services on Azure}

\input{abstract}

\ccsdesc[500]{Information systems~Autonomous database administration}
\ccsdesc[300]{Computer systems organization~Cloud computing}
\keywords{cloud, data service, autonomous data service, self-driving data service, artificial intelligence (AI), machine learning (ML), data science (DS), ML for system, cloud infrastructure, query engine}

\maketitle


\section{Introduction}

Modern cloud has made access to various data processing systems easier than ever. Today, Azure---Microsoft's public cloud offering---provides a large range of data services to customers, including SQL databases (e.g. SQL Server, PostgreSQL, and MySQL), NoSQL databases (e.g. CosmosDB), analytics (e.g. Synapse SQL and Synapse Spark), big data (e.g. Apache Kafka and Apache Storm), and BI (e.g. PowerBI Analysis Service). While cloud providers benefit from the economy of scale, migration to cloud also brings with it complexity. 
As the number and the complexity of data services continue to grow, cloud providers are facing increasing difficulties in managing all aspects of a service, such as resource provisioning, scheduling, query optimization, query execution and service tuning, while still satisfying customer SLAs and reducing operational expenses.
For cloud users, on the other hand, it is non-trivial to extract the maximum benefit from these data services, with each service exposing many configurations and performance knobs to tune. The recent trend of \textit{serverless computing} seeks to relieve users from the burden of choice. However, this product line simply transfers the problem from cloud users back to cloud providers. Automating data services thus is an integral part of the cloud to operate at scale.

While the cloud brings with it complexity, it presents massive opportunities. We have never before had access to such detailed workload traces and system telemetries, collected across millions of users and applications.
More instrumentation is continuously added to the cloud for better tracing and monitoring. The combination of the recent advances in data science (DS) and machine learning (ML), sophisticated telemetry, and shortage of data experts make now an ideal time for the development and adoption of \textit{autonomous data services}. Prior research on self-adaptive~\cite{selfadapt}, self-tuning~\cite{selftune}, and self-managing~\cite{selfmanage} databases has been ongoing for decades, but it is only with the advent of cloud technology that the practicality of autonomous data services has emerged. 
Oracle announced the ``World’s First Self-Driving Database" in 2017~\cite{oracleselfdrive} suggesting that ML will replace DBA, followed by many efforts on autonomous databases from industry and academia~\cite{ottertune, Pavlocidr17}. 
Furthermore, there is a wealth of research focused on utilizing ML to improve or substitute various components of database engines, such as the cardinality estimator, cost model, query planner, and indexer~\cite{learnedCE,learnedqo,learnedindex}. 
We are witnessing an explosion of DS/ML-for-Systems innovations applied in the area of autonomous databases.

\edit{The vision described in the paper is the distillation of multiple research efforts led by applied researchers and data scientists from the Gray Systems Lab (GSL)~\cite{gsl}, in close collaboration with engineers from various departments within Azure Data. The set of research initiatives seek to enhance and automate different facets of Azure Data services, which
have yielded significant COGS (cost of goods sold) saving for Azure. 
In this paper, we present}
 our perspectives on the development of autonomous cloud data services, including the challenges involved, the progress we have made, the lessons we have learned, the future directions we are pursuing, and the outstanding questions that require further investigation.

\section{Our Viewpoints}

We first present our viewpoints on building autonomous data services in the cloud and explain the rationals behind them.


\viewpoint{Viewpoint 1}{The economic scale that has driven the adoption of cloud technology has also necessitated the development of autonomous data services. However, we contend that true autonomous data services can only be achieved in the cloud, meaning that the cloud is a necessary precondition for the attainment of autonomy in data services.}

Gaining knowledge from past experiences, which may span multiple users and applications, is a critical step towards achieving autonomy for data services. The cloud platform provides extensive visibility into a vast array of system metrics and workloads from numerous users and applications over time. Due to its expansive range of services and customer base, the cloud amortize the cost of advanced quality-of-service (QoS) features, making it more financially viable to invest in ML-based solutions. Although a 1\% improvement in on-premise systems for an individual customer may seem insignificant, when applied across millions of cloud users, it can have a substantial impact. Additionally, as customer workloads continue to evolve, the learning process must adapt accordingly, which the cloud facilitates through the rapid deployment of updates, often without requiring end-user involvement.


\viewpoint{Viewpoint 2}{Autonomy spans all layers of data services: cloud infrastructure layer, query engine layer, and service layer.}

Developing an autonomous data service on the cloud requires leveraging ML to improve or replace more than just the work of DBAs and individual engine components. To illustrate, let us consider the life cycle of a query on a serverless big data service. These services offer a range of adjustable knobs that can impact system performance. For example, Spark requires the user to specify the number of executors and resources (cores and memory) allocated to them. Since this is a serverless service, all such decisions must be made automatically at the \textit{service layer}. Prior to running the query, the service must be operational. If this is the first time the customer has used the service, then VM or container resources must be provisioned at the \textit{cloud infrastructure layer}. This raises a number of questions, such as which VM or container SKU to select, what the software configuration should be, and whether proactive resource provisioning is necessary to meet SLA for the customer. If so, what SKUs should be proactively provisioned? Once the service is running, the query must be optimized using an accurate cardinality, the correct cost model, and a reliable query planner. The query is then executed efficiently, potentially utilizing indexes and materialized views recommended based on the workloads. It is evident that all layers of the cloud stack, including the cloud infrastructure layer, query engine layer, and service layer, as well as the interactions among them, must be taken into consideration when creating autonomous data services. This level of complexity can be daunting for many institutions.

\viewpoint{Viewpoint 3}{The objectives of autonomous data services are: improving ease of use, optimizing performance, reducing costs, and maintaining data privacy.}



Autonomy is not the ultimate goal of cloud services, but rather a means to achieving simpler, faster, and more cost-effective services for users, while also prioritizing data privacy. Simplicity or ease of use is an important aspect, whereby users should not have to worry about resource allocation, query optimization, or excessive configuration and tuning decisions to use a data service. Achieving optimized performance is a shared goal among cloud users and providers. To meet this goal, the cloud infrastructure layer needs to offer fast and intelligent resource provisioning and scheduling, queries must be optimized and executed efficiently, and services should be appropriately tuned. Cost savings benefit both users and providers. Interestingly, as we will show later, many times performance and cost saving can be achieved simultaneously, but sometimes they are at odds with each other and requires a trade-off between these two goals.
Finally, preserving privacy must be a fundamental requirement when pursuing the other three objectives.

\section{The Vantage Point}
\edit{
As an applied research organization under Azure Data, GSL is situated at the intersection of research and product development. Our viewpoints on building autonomous cloud data services are therefore shaped by the rich research and our first-hand experience working with various product groups in Azure Data.
}

\textbf{Research in Autonomous Data Services.} Our perspectives and progress in creating autonomous data services are built upon a strong foundation of research from the academic community. In the interest of space, we focus on the major trends and provide examples of influential works, although a comprehensive survey of research in this area is beyond the scope of this discussion. As pointed out by~\cite{pavlovldb21}, the work on autonomous data systems dates back more than four decades, with the development of self-adaptive databases~\cite{selfadapt}. In the early 2000s, projects such as Microsoft's AutoAdmin~\cite{autoadmin} and IBM's DB2 design advisor~\cite{db2advisor} marked the era of self-tuning databases~\cite{selftune}. Oracle subsequently introduced the self-managing database~\cite{selfmanage}. While these earlier works aimed to alleviate various administrative tasks for DBAs, such as memory allocation, index recommendations, and materialized views, they did not utilize DS\&ML techniques. However, the cloud and the advancements in DS\&ML technologies have accelerated progress towards autonomous data services. In 2017, Oracle announced the ``World's First Self-Driving Database''~\cite{oracleselfdrive}. In academia, several efforts have focused on autonomous or self-driving databases~\cite{Pavlocidr17, ottertune}, which aim to automatically tune database configurations or optimize databases for predicted future workloads. In a related line of research, many studies have applied ML to improve database engine components, often referred to as learned components. These include learned indexes~\cite{learnedindex}, learned cardinality estimation~\cite{learnedCE}, learned query optimizer~\cite{learnedqo}, and learned checkpoint~\cite{phoebe}. All of these efforts occur either inside the database engine or on top of the engine in the service layer. The efforts in optimized infrastructure support~\cite{dassigmod16, dastods13} for cloud data services are fewer in comparison.

\textbf{First-Hand Product Experience.} 
Over the years, we have worked on a large number of ML-for-Systems projects, and successfully delivered many new or improved features in making the corresponding Azure data services more autonomous. Horizontally, we have worked on Cosmos~\cite{cosmos} (an internal cloud data service in Microsoft), Azure SQL Database~\cite{sqldb}, Azure Synapse SQL~\cite{synapsedw}, Synapse Spark~\cite{spark}, HDInsight~\cite{hdinsight}, etc. Vertically, our work has touched all three layers of data services. We have proposed novel techniques as well as adapted existing state-of-the-art research ideas to address practical concerns such as explainability, debuggability, and cost management. In Section~\ref{sec:progress}, we will showcase some of these projects. 



\section{Challenges and Progress}\label{sec:progress}

In this section, we discuss the challenges in automating each layer of the cloud stack and report on the progress we have achieved.


\subsection{Cloud Infrastructure Layer}

The cloud infrastructure manages all hardware and software resources for the life cycle of data services. Significant technical and research efforts have been made to enhance it, including resource provisioning~\cite{hadary2020protean}, job scheduling~\cite{boutin2014apollo,grandl2016graphene}, container imaging~\cite{memcached}, and autoscaling~\cite{tirmazi2020borg,floratou2017dhalion}. However, these components heavily depend on the manual adjustments by experts in the field, with fixed parameters dispersed throughout the code base in configuration files. With the emergence of advanced analytical tools and abundant telemetry data, new opportunities for automation arises.
Our solutions in this layer were built based on our findings on the predictability of system behaviors and user behaviors.


\textbf{Modeling system behaviors based on domain knowledge and system metrics.} 
Training models for autonomous data services requires a substantial amount of data from various system tunables. While existing observational data may suffice in scenarios with inherent volatility~\cite{zhu2021kea}, additional data is often necessary. However, gathering such data through rounds of trials on the production infrastructure is impractical due to potential service disruptions. As a result, we must either devise ways to minimize the number of experiment runs or gather system metrics and use ML to ``emulate'' system dynamics. In both cases, domain knowledge is crucial to comprehend the causal links among different components and establish trustworthy models of the complex system.



\begin{figure}
 	\centering
 		\subfloat[CPU vs Task Exec Time]
 		{\includegraphics[width=0.48\columnwidth]{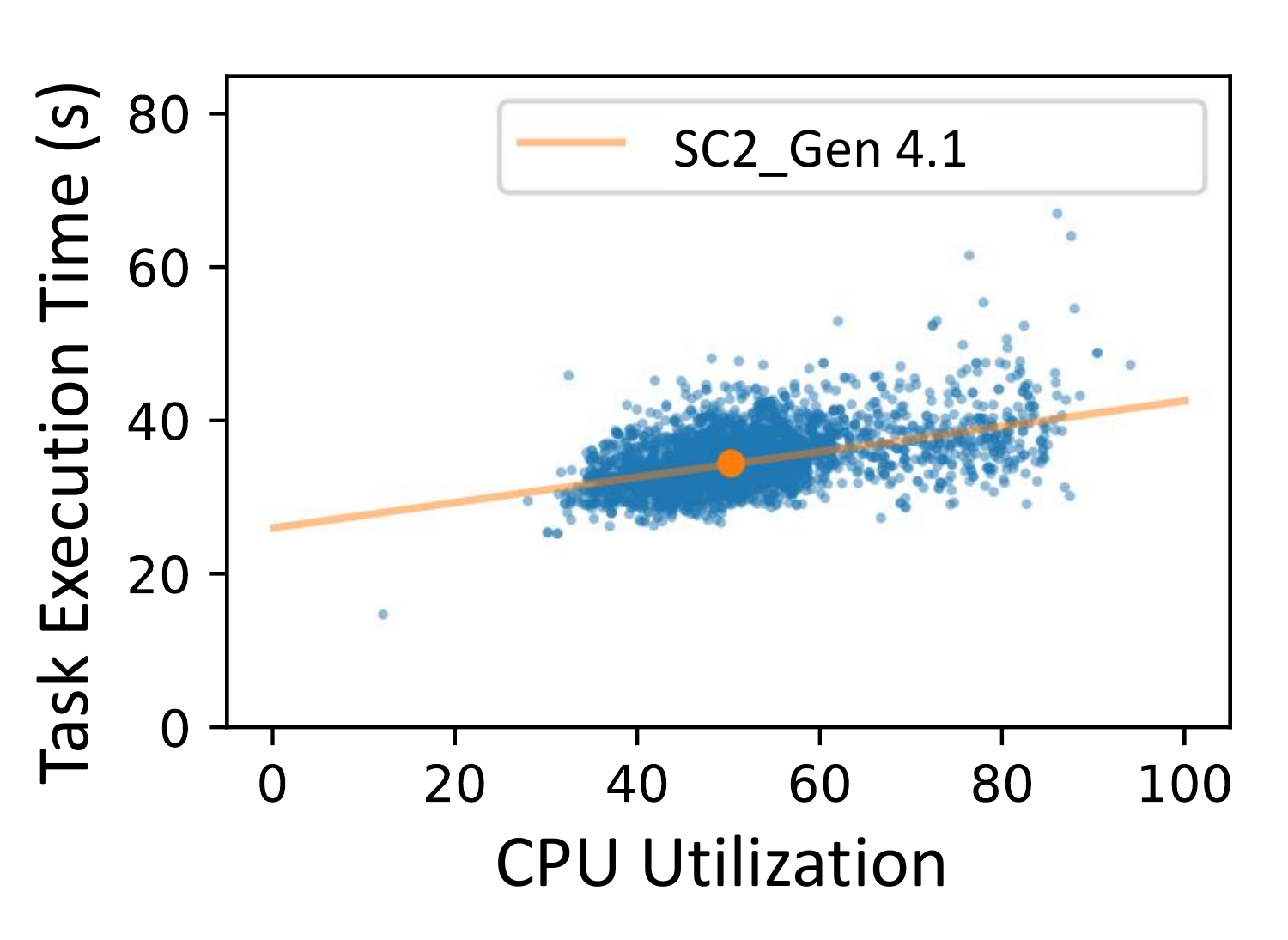}\label{fig:median-vs-runtime}\vspace{-0.2cm}}
 		\subfloat[CPU vs Runinng Containers]
 		{\includegraphics[width=0.51\columnwidth]{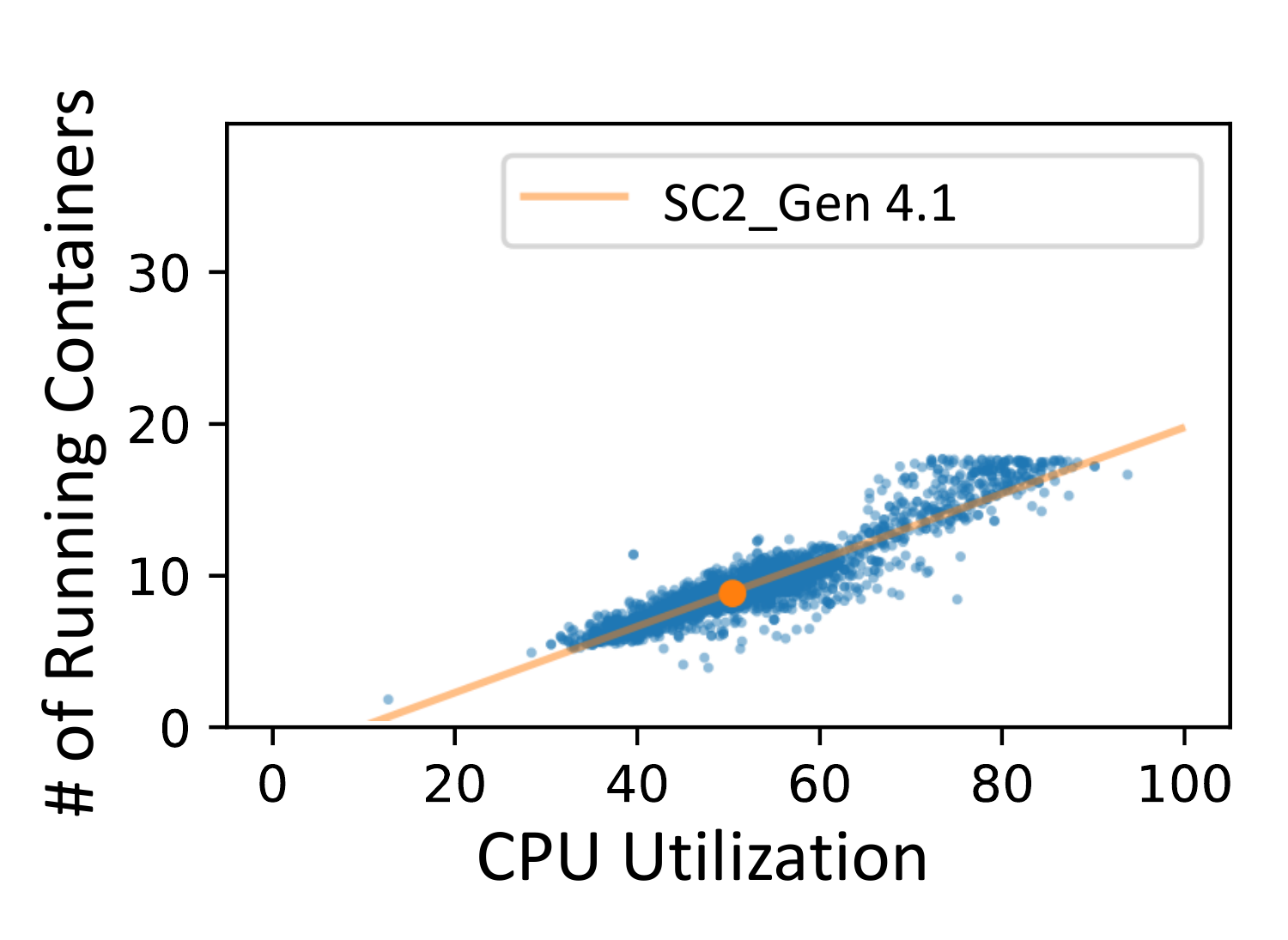}\label{fig:cov-as-predictor}\vspace{-0.2cm}}
 		\vspace{-0.3cm}
 	\caption{Models to predict machine behavior~\cite{zhu2021kea}}
 	\label{fig:kea}
 \end{figure}

As an example, in~\cite{zhu2021kea}, we employed multiple linear models to predict machine behavior, such as CPU utilization versus task execution time or the number of running containers (see Figure~\ref{fig:kea}). These models were then integrated into an optimizer to balance workloads by tuning Cosmos scheduler configurations, such as the maximum running containers for each SKU. Similar methods were used to determine the hardware/software configuration, such as RAM/SSD size and the mapping of logical drives to physical media, and to set power limits on Cosmos racks. For Azure Synapse Spark~\cite{spark}, we developed a simulator to mimic the cluster initialization process and derived the optimal policy for sending requests, reducing its tail latency. As another example, by using ML to predict the throughput and latency of benchmark workloads on VMs with various kernel parameters, developed on MLOS~\cite{curino2020mlos}, we refined the parameters of the Azure VM that runs Redis workloads.

\begin{figure}[tp]
	\centering
	\includegraphics[width=0.6\columnwidth]{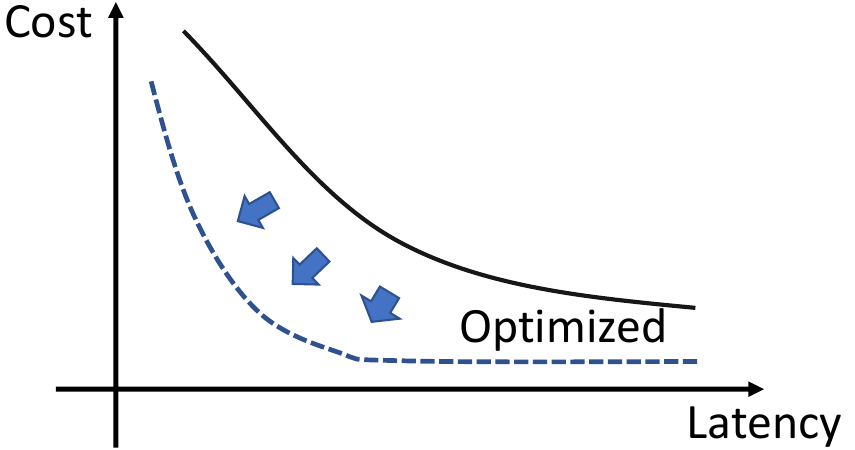}
	\caption{Pareto curve depicting the trade-offs between the QoS (x-axis) and the cost (y-axis)
	}\label{fig:pp}
\end{figure}

\textbf{Modeling user behaviors for better trade-offs between quality of service (QoS) and cost.}
Cloud operators face a continuous challenge in managing resources, striking a balance between QoS, such as low latency, and operational costs. To fulfill customer SLAs, cloud operators often need to proactively provision resources, which can lead to additional expenses. This interdependence is illustrated by the Pareto curve in Figure~\ref{fig:pp}. By utilizing ML, these trade-offs can be measured, and the Pareto curve can be globally optimized. In~\cite{poppe2022moneyball}, we demonstrated that 77\% of Azure SQL Database Serverless usage is predictable and used ML forecasts to pause/resume databases proactively. Another instance is proactive cluster provisioning based on expected user cluster creation demand to reduce wait time for cluster initialization on Azure Synapse Spark, optimizing both COGS and performance.

\subsection{Query Engine Layer}

Despite the significant amount of research conducted on utilizing ML techniques to enhance or replace parts of the query engine~\cite{learnedCE, learnedindex, learnedqo}, there is still reluctance within the industry to apply these advanced methods to actual production systems. This reluctance can be attributed to several factors. Firstly, real production systems are often more intricate than the academic prototypes used in research papers. For instance, the start-of-the-art learned optimizer, Bao~\cite{learnedqo}, which provides rule hints to steer the optimizer towards better plans, only takes into account 48 rule configurations, \edit{whereas the SCOPE query engine~\cite{cosmos} used in Cosmos has 256 rules in the query optimizer, which leads to $2^{256}$ rule configurations that need to be considered}. Secondly, while sophisticated ML algorithms have demonstrated superior performance over current engine components, production engineers prioritize the interpretability and debuggability of the models, as every new feature introduced may generate new incident tickets for on-call engineers to resolve. Thirdly, workload patterns change over time due to data or concept drift, and regression is a genuine concern. Finally, the cost of training, especially for deep neural networks, becomes as another obstacle to their adoption.


This subsection outlines our efforts to automate various aspects of query engines in production environments and address the aforementioned challenges. The fundamental principle underlying our work is to \textbf{learn from the past to improve the future}. Our work is based on the observation that in actual production workloads, \textbf{queries and jobs are often recurrent and similar}. For instance, in SCOPE, over 60\% of jobs are recurring (involving periodic runs of scripts with the same operations but different predicate values~\cite{knobadvisor}), and nearly 40\% of daily jobs share common subexpressions with at least one other job~\cite{cloudview}. This highlights the potential benefits of insights learned from the past workloads to improve the efficiency of future workloads.



\textbf{Workload Analysis.} To automate query engines, we start from workload analysis~\cite{peregrin}. There are several pieces of information that are crucial for learning: meta data, query logs, and run time statistics (such as execution time and actual cardinality). However, these data sources are frequently dispersed in different locations. Consequently, our first step is to combine this information. To facilitate various applications of the workload data, queries or subexpressions of queries are categorized into templates based on their recurrence and similarity, and the dependencies of queries/jobs (where the output of one job serves as the input of another) in pipelines are captured~\cite{peregrin}. Furthermore, workloads evolve over time, and as such, we also learn the evolving nature of the historical workloads to forecast future workloads.


\textbf{Query Optimization.} The optimizer serves as the brain of a query engine, and decades of research and development have been invested in improving this component for any data engine on Azure. Our guiding principle is to \textbf{minimize changes to the existing optimizer and supplement it with learned components}. Specifically, we \textit{externalize} the learned components and add simple extensions to the optimizer to leverage these external services. For cardinality estimation, we utilize the templates generated by workload analysis and train per-template micromodels~\cite{cardlearner}. We reduce the number of micromodels by retaining only those that would actually improve performance. Consequently, the optimizer can employ more precise cardinalities for queries or subexpressions with corresponding models while reverting to the default cardinalities for others. We adopt the same micromodel approach for learned cost models~\cite{costmodel} and introduce a meta ensemble model that corrects and combines predictions from individual models to increase coverage. To enhance optimizer plans using rule hints, we have made notable progress in applying state-of-the-art research ideas from Bao~\cite{learnedqo} to production settings. However, we had to make significant adjustments for the production system, including limiting steering to small incremental steps for better interpretability and debuggability, minimizing pre-production experimentation costs using a contextual bandit model, and guarding against regression with a validation model~\cite{knobadvisor1, knobadvisor}.

\textbf{Query Execution.} In query engines of big data services like Cosmos and Spark, a job is compiled into a Direct Acyclic Graph (DAG) of stages  that are executed in parallel. In the case of Cosmos, we have observed an increase in the job complexity over the years, with some jobs containing thousands of stages~\cite{phoebe}. 
During runtime, these large jobs can lead to machine hotspots that run out of local temporary storage space, longer restarting times in case of failures, and suboptimal performance due to compounding errors from poor optimizer estimates. In~\cite{phoebe}, we trained models to estimate the execution time, output size, and start/end time of each stage taking into account of the inter-stage dependency, then applied a linear programming algorithm to introduce checkpoint ``cut(s)'' of the query DAG. With this checkpoint optimizer, we were able to free the temporary storage on hotspots by more than $70\%$ and restart failed jobs $68\%$ faster on average with minimal impact on Cosmos performance.

\textbf{Computation Reuse.} With the large portion of recurrent and overlapping queries observed in real production workloads, there is a great opportunity to reuse past computations for future queries. CloudViews~\cite{productionCloudViews, sparkcruise} was developed to detect and reuse common computations on Cosmos and Spark. It relies on a lightweight subexpression hash, called a signature, for scalable materialized view selection and efficient view matching. Deployed on Cosmos, we have observed $34\%$ improvement on the accumulative job latency, and $37\%$ reduced total processing time~\cite{productionCloudViews}. We have worked on improvements of CloudViews on several fronts, including extending the reuse from the \textit{syntactically} equivalent subexpressions detected by the signatures to \textit{semantically} equivalent and contained subexpressions while still maintaining the efficiency and scalability of the detection process, as well as enabling a query to \textit{partially} take advantage of a view with the remaining results computed on the base tables.

\textbf{Pipeline Optimization.} Production workloads not only have many recurrent queries, but also many recurrent query pipelines, where queries are interconnected by their outputs and inputs. For example, $70\%$ of daily SCOPE jobs have inter-job dependencies. We \edit{analyzed the interdependency to facilitate job scheduling~\cite{wing} and }developed a pipeline optimizer to optimize these recurrent pipelines~\cite{pipemizer}, including collecting pipeline-aware statistics and pushing common subexpressions across consumer jobs to their producer job.

\subsection{Service Layer} \label{sssec:4.3}

DS\&ML solutions impact how customers engage with a system at the service level. The primary goal of the autonomous cloud services is to automate as many customer-facing decisions and options as possible while also providing highly customizable solutions. DS\&ML tools allow for the automation of various decisions by studying customer and application profiles. We can develop models with different levels of granularity: 1) a global model that is broad but may not be precise, 2) a segment model that groups similar customers or applications and shares insights within the group, and 3) an individual model for each customer or application that requires sufficient data observations.

\textbf{Individual models are more accurate when there is enough data.} To automate the scheduling of backups for PostgreSQL and MySQL servers, we used ML models to forecast user load for each specific server~\cite{poppe2020seagull}. The system identifies low load windows with 99\% accuracy, and the solution has been deployed for tens of thousands of PostgreSQL and MySQL servers across all Azure regions.

\textbf{Segment models or global models are deployed jointly to transfer learning across customers/applications.} To automate the SKU suggestion for migrating from on-premise SQL Server to the cloud, we proposed a profiling model that compares new customers to existing segments of Azure customers. 
This enables new customers to benefit from the decisions made by customers with similar characteristics.
We achieved a recommendation accuracy of over 95\% by combining the segment-wise knowledge with a per-customer price-performance curve that offers a customized rank of all SKU options~\cite{doppler}.
Another example involves auto-tuning configurations for Spark, built on top of the resource usage predictor~\cite{sen2020autotoken}. We use iterative tuning algorithms to replace the manual process for customers. We start with a global model trained using data from multiple benchmark queries. While the global model may not be highly accurate, it serves as a reasonable starting point and is fine-tuned for each application as more observational data becomes available.

\section{Lessons Learned}
\label{sec:lessons}





Given our experience developing and integrating DS\&ML solutions at the cloud infrastructure, query engine and data service layers for various cloud services across Microsoft (e.g., SCOPE~\cite{cosmos}, Synapse DW~\cite{synapsedw} as well as Spark~\cite{spark, hdinsight}), common patterns arose across our engagements. 
Here we list some key lessons that we believe underlie our production successes and the speed at which we generate value for our product partners.

\viewpoint{Insight 1}{Simplicity rules.}

The common pattern across all our engagements is that simple heuristics tend to overrule ML and simple ML models, like linear models and tree-based models, tend to overrule complex deep learning models. This is particularly true for new engagements, with teams that have yet to adopt ML within production. For example,  in~\cite{poppe2020seagull}, for PostgreSQL or MySQL servers that follow a stable daily or a weekly pattern, a simple heuristic that predicts the load of a server based on that of the previous day was already sufficient to generate 96\% accuracy.
There are many projects in which a linear regression was most appropriate~\cite{sen2020autotoken,zhu2021kea}.
Simplicity helps with:

\mypar{\textit{Cost}} While there is growing consensus on the positive impact of ML in automating and optimizing cloud services, production deployments have to evaluate trade-offs with the increase in COGS to enable it. Consequently, algorithms are selected, not only by performance, but also by other factors, like training and inference cost, dependency on specialized hardware (GPU, FPGAs), etc..  

\mypar{\textit{Scalability}} For production systems, we need the metrics to seamlessly scale in training time, re-train frequency and data/parameter handling. Most sophisticated machine learning techniques do not satisfy this fundamental requirement, e.g., reinforcement learning requires substantial training data before outperforming traditional approaches. Moreover, when inference is on the critical path (which impacts the customer-experienced latency), latency becomes crucial for the design of the infrastructure which prunes the solutions space considerably.

\mypar{\textit{Manageability}}
Manageability is important in two dimensions --- debuggability and upgrades/rollbacks. ML models are sometimes notorious for their difficulty in debugging. In a production environment, when encountering regression, a complex data lineage \textit{across a multitude of systems and language} is needed for a close investigation from data ingestion to model (deployed) inference\cite{vamsa}. Debuggability needs to be well-supported with tracking/versioning through MLOps~\cite{sysforml} for continuous integration. 

\mypar{\textit{Explainability}} For customer-facing solutions, the expectation is that the reasoning of a choice made under the hood by any algorithm has a succinct and ideally intuitive rationale. In this sense, an explainable solution, which in turn translates to simplicity such as~\cite{doppler}, is very much preferable, while also improving the manageability as mentioned before. 

\viewpoint{Insight 2}{One size does not fit all.}

One global (macro) model that functions reasonably well for all scenarios can typically be traded off against several specific (micro) models that are tailored for individual customers, as discussed in Section~\ref{sssec:4.3}.
Identifying and crafting a single global model is generally difficult, as data heterogeneity necessitates considerable feature construction and model hyperparameter tuning for optimal performance.
Micro models, however, go against simplicity due to the challenges in managing the large number of models. A happy middle ground can be achieved by identifying natural ways to stratify the data, and building micro models for each cluster as done in the SKU recommendation framework~\cite{doppler} that recommends right-sized Azure SQL SKU to migrate on-premise databases.

\viewpoint{Insight 3}{Feedback loop is indispensable.}


It is universally accepted that all ML solutions undergo extensive testing before being deployed into production, including back-testing, flighting~\cite{zhu2021kea} or A/B testing (potentially with a smaller group). 
The dynamic nature of cloud data services, however, necessitates ongoing improvement of even "well-tested" solutions in order to maintain performance, which leads to requirements for (1) a thorough monitoring system to spot potential changes in real-time, continually assess, and initiate fine-tuning of the model, and (2) a rollback mechanism that reacts fast and avoids regression.

\section{Future Directions}

In this section, we discuss some of the future directions that we are currently pursuing while also highlighting the challenges. 

\viewpoint{Direction 1}{Reuse, reuse and reuse!}

Despite the differences between distinct data services on Azure, they all face a set of similar issues. For example, at the infrastructure level, many services need efficient cluster provisioning and auto-scaling. At the engine level, many require improvement in cardinality estimation, query planning, and computation reuse. At the service level, auto-tuning is highly sought after for many services. Working on similar issues with multiple Azure data services over time, we came to the realization that \textbf{a common reusable solution is highly desirable to efficiently leverage the similar technologies and software artifacts among multiple services}.

However, reality presents a lot of challenges to reusability. Distinct services collect different service-specific telemetries and workload traces, store them in different places (e.g., Kusto\cite{kusto}, SQL server, etc.), and have different preferences on the infrastructure for model deployment (e.g. AML\cite{aml}, Synapse ML\cite{synapseml}, etc.). 

So, can we reach the holy grail of reusable ML solutions? Although we don't have a complete answer, we can perhaps try to tackle this problem at different granularities of reusability.

\textbf{Function Level Reuse.} In the finest granularity, we can reuse pieces of code modules that implement specific functions, e.g., time series analysis of OS performance counter data. Our proposal is to create a \emph{AlgorithmStore (analogous to a GitHub for models)}, which is a project gallery with predefined algorithm templates. The previously developed algorithm can be discovered and adapted to address new scenarios quickly. For this type of algorithm catalog, it is required to have: (1) an easy search interface to discover similar pre-existing solutions; (2) good API design to support extensibility and customizations; (3) clean modularized functions; (4) significant coverage of common use cases; (5) code quality to allow robust reuse; and (6) better documentation.

\textbf{Component Level Reuse.} At the component level, the question of reusability pertains to whether we can establish a shared infrastructure that supports similar or related system components across various data services. For example, can we develop a common infrastructure that facilitates auto-scaling for all services or query optimization for all data engines? This task becomes increasingly difficult due to the aforementioned differences among distinct services. Nevertheless, we have made some strides on this front. The Peregrine workload optimization platform~\cite{peregrin} represents a common infrastructure for a set of related engine problems, such as cardinality estimation, cost models, and computation reuse. It has been implemented for both Cosmos and Spark. Peregrine consists of an engine-agnostic workload representation, workload categorization based on patterns, and a workload feedback mechanism that enables query engines to respond to workload feedback.

\textbf{System-for-ML Support Level Reuse.} At the highest level of granularity, all ML-for-Systems projects require System-for-ML support, from data ingestion, featurization, model training and tuning, model deployment, to model tracking. In GSL, we have a large collection of System-for-ML projects towards building such a common infrastructure. A summary and vision of our efforts in this area is provided in~\cite{sysforml}. 

\viewpoint{Direction 2}{Standardization.}

Standardization is critical for developing reusable infrastructure across data services. It begins with telemetry. In addition to structuring the collected telemetry similarly across platforms and data services through initiatives such as OpenTelemetry~\cite{opentelemtry}, we are also exploring the use of semantic information from telemetry to enhance reusability across platforms and services (e.g. CPU utilization metrics on Windows and Linux VMs possess the same meaning even though they may have different names). At the query engine level, we require standardization for representing workloads and query plans. We have made some initial efforts on an engine-agnostic workload representation as part of the Peregrine workload optimization platform~\cite{peregrin}. We are now exploring the use of cross-language query plan specification, such as Substrait~\cite{substrait}, as a standard plan representation across our engines. To simplify the reuse of models for deployment within a common infrastructure, we also adopt standard representations for ML models, such as ONNX~\cite{onnx}. Furthermore, we package an ML model (along with any additional required code and libraries) into a standard generic container that can be efficiently reused across systems~\cite{cage}, making it portable across all of our model-serving capabilities at Microsoft.

\viewpoint{Direction 3}{Optimization across components jointly.}

In many projects, the primary focus is typically on optimizing a single component of the entire system since it is owned by a specific product team. For example, VM provisioning is owned by the cluster service team, while cardinality estimates are owned by the query optimizer team, and so on. However, sequentially optimizing each individual component is unlikely to yield optimal overall performance. Conversely, for a complex cloud service, especially at scale, it is impractical to create a massive optimization problem that simultaneously optimizes all components while accurately capturing interactions across different components. 
Ongoing efforts continue to jointly optimize a selection of components and synchronize the deployment of changes so that the observational data reflects the latest deployed configuration. This approach enables us to focus on optimizing related components that work together in a coordinated manner. By improving the joint optimization of these components, we can improve the overall system performance.

\viewpoint{Direction 4}{Responsible AI (RAI)}

ML cannot be applied without risks~\cite{jindalquery}, e.g., over-indexing on a particular customer or workload, and bias is an inherent problem that we continually encounter. We introduce guardrails to protect customers from expensive solutions and from performance regressions, and we regularly check that our ML-driven decisions serve all customers fairly. We have a responsibility to ensure that customers, big or small, do not get marginalized from autonomous decisions. 

At Microsoft, we are operationalizing the Responsible AI (RAI) at scale to protect privacy and security, improve fairness, inclusiveness, reliability, safety, transparency, and accountability. 
For the ML-related projects, we perform a comprehensive RAI assessment which is for now a manual and prolonged process by domain experts. Several automation tools were developed (e.g.,~\cite{holstein2019improving}), however, ad-hoc solutions are still required for many cases.

\section{Conclusion and Call to Action}

We are living in fascinating and rapidly evolving times where technology is advancing at a breakneck pace. 
Cloud and AI are among the most transformative technologies of our era. The intersection of these two revolutionary technologies can be witnessed in the progress made towards autonomous data services on cloud. In this paper, we showcased some of our progress in automating data services on Azure. However, challenges remain to be overcome as highlighted in the previous section. We believe that the database community has a vital role to play in shaping the future of cloud data services. We welcome other researchers to join us in this exciting journey.

\section*{Acknowledgements}

We thank past team members, interns, and MAIDAP collaborators for their contribution to our progress.



\balance
\bibliographystyle{ACM-Reference-Format}
\bibliography{bibliography}

\end{document}

%% file: abstract.tex
\begin{abstract}

Modern cloud has turned data services into easily accessible commodities. With just a few clicks, users are now able to access a catalog of data processing systems for a wide range of tasks.
However, the cloud brings in both complexity and opportunity.  
While cloud users can quickly start an application by using various data services, it can be difficult to configure and optimize these services to gain the most value from them.
For cloud providers, managing every aspect of an ever-increasing set of data services, while meeting customer SLAs and minimizing operational cost is becoming more challenging. 
Cloud technology enables the collection of significant amounts of workload traces and system telemetry. With the progress in data science (DS) and machine learning (ML), it is feasible and desirable to utilize a data-driven, ML-based approach to automate various aspects of data services, resulting in the creation of \textit{autonomous data services}. This paper presents our perspectives and insights on creating autonomous data services on Azure. It also covers the future endeavors we plan to undertake and unresolved issues that still need attention.
\end{abstract}